# COMPARATIVE ANALYSIS OF QUALITY OF SERVICE SCHEDULING CLASSES IN MOBILE AD-HOC NETWORKS

Thulani Phakathi, Bukohwo Michael Esiefarienrhe
and Francis Lugayizi

Department of Computer Science, North-West University,
Mafikeng, South Africa

*ABSTRACT*

*Quality of Service (QoS) is now regarded as a requirement for all networks in managing resources like bandwidth and avoidance of network impairments like packet loss, jitter, and delay. Media transfer or streaming would be virtually impossible if QoS parameters were not used even if the streaming protocols were perfectly designed. QoS Scheduling classes help in network traffic optimization and the priority management of packets. This paper presents an analysis of QoS scheduling classes using video traffic in a MANET. The main objective was to identify a scheduling class that provides better QoS for video streaming. A simulation was conducted using NetSim and results were analyzed according to throughput, jitter, and delay. The overall results showed that extended real-time Polling Service (ertPS) outperformed the other classes. ertPS has hybrid features of both real-time Polling Service (rtPS) and Unsolicited Grant Service(UGS) hence the enhanced performance. It is recommended that ertPS scheduling class should be used in MANET where QoS consideration is utmost particularly in multimedia streaming applications.*

## 1. INTRODUCTION

Quality of Service (QoS) is a widely used term associated with the measurement of the overall performance of a network service in multimedia applications. It is a necessary function within MANETs to ensure an end to end quality in terms of throughput, jitter, bandwidth, or delay. One of the biggest challenges in MANETs is the design of a secure and efficient routing protocol that will guarantee an acceptable level of QoS during the routing process as MANETs communicate only when they are in the range of each other or near a base station. There is no central controlling device. These design characteristics make it difficult to ensure QoS in the network.

The support for QoS [1] constantly requires link-state information to be forever present i.e. bandwidth, loss rate, error rate, and delay. Supporting the mentioned requirements is often a huge task as the quality of the ad-hoc link can abruptly change because nodes are mobile and operate in dynamic environments. Earlier approaches to QoS were based on the virtual circuit model which meant that there should be fixed connection management before communication so that for the duration of that session there would be a guarantee of route and reservation between the source node and the destination node. This was a great approach but it lacked the flexibility needed to dynamically adapt to mobile ad-hoc networks where the path and reservation needed to respond dynamically and in real-time to the ever-changing topology and resource needs [2]. One of the threats to QoS support is the maintenance of an acceptable service level. By default, QoS support in MANETs is linked to the routing protocol's performance because the flow between the source nodes to the destination is not always straightforward. There is a high chance that within the lifetime of an on-going session, the flow gets rerouted and this happens frequently. When this happens, there is also a change in resource provisioning and needs as the flow would be now on a new path. The QoS agreement from the initial path may not be valid and so would be the assumption that the route and reservation remain fixed in that duration of a session. This is the reason why many researchers have been constantly coming up with mechanisms and frameworks that can support QoS in MANETs adaptively and can respond to its intrinsic behavior [1].

In MANETS, it is difficult to guarantee QoS than traditional networks because the wireless bandwidth is shared amongst nodes and the network topology is forever changing as nodes are in transit. It is for this reason that QoS provisioning can only be achieved through extensive collaboration among nodes for route establishment and securing the available network resources. The provision of QoS in a network can be classified into two namely soft and hard QoS approaches. In the soft QoS approach, the QoS requirements are not guaranteed for the whole session. This may be due to insufficient available network resources. Hard QoS approaches ensures the availability of network resources to meet all the QoS requirements of a connection and such requirements can be sustained for the entire duration of the session. QoS Provisioning improves end to end performance of nodes in heavily congested network scenarios through QoS aware routing, resource reservation, admission control, scheduling, and traffic analysis. Some of the other reasons why QoS provisioning is still a challenge in MANETs is because of limited resource availability, error-prone channel, lack of central controller, hidden terminal problem, insecure medium and dynamic network topology. This makes it hard to achieve and maintain end to end QoS. The most crucial system components to QoS provisioning in MANETs are network resources and their availability to process application data. There is no standardized mechanism to guarantee absolute QoS in MANETs but only some level of QoS can be achieved through different methods.

This work is arranged as follows: Section 2 discusses mechanisms that have been implemented in literature according to their various categories, section 3 presents the proposed framework in detail including the sinkhole attack implementation and the counterattack intrusion detection system. Section 4 presents the results and discussion and lastly, section 5 gives the conclusion and suggestion for future work.

## 1.1. IEEE 802.16 & QoS Scheduling Approaches

The IEEE 802.16e is an amended version of the IEEE 802.16 that was initially approved for targeted fixed applications. The amendment was enhanced for adding mobility support for nomadic and mobile applications. The new additions introduced modifications in the physical layer from OFDM to scalable OFDMA[3]. There were additional modifications within the MAC layer for resource management, roaming, security, and handoff. Additional enhancements in the MAC include sleep/idle-mode for mobile nodes, power-saving classes, paging, locating, and defining messages for handover procedures. The IEEE 802.16e specification adds a new scheduling service called extended real-time Polling Service (ertPS), which combines the efficiency of Unsolicited Grant Service (UGS) and real-time Polling Service (rtPS). It allows unsolicited bandwidth grants like UGS, but with dynamic size like rtPS. This yields a services class supporting real-time service flows with variable size data packets, suitable for Voice over IP (VoIP) with silence suppression[4].

**1.1.1. QoS Scheduling Classes**

The scheduling process determines the order in which packets in a queue should be processed. Priority scheduling involves using an algorithm that allows the router to fix the priority level for the packets that would be coming from different sources and directions. Higher priority packets are processed first and sent out.

### A. Unsolicited Grant Service (UGS)

The UGS scheduling service type is designed to support real-time data streams consisting of fixed-sized data packets issued at periodic intervals, for example, Pulse Code Modulation(PCM) pone signals and Voice over IP without silence suppression. The base station provides data grants at fixed periodic intervals to reduce the latency and overhead of the subscriber stations. This class is used for high priority packets[5].

### B. Real-time Policing Service (rtPS)

The rtPS scheduling service type is designed to support real-time data streams consisting of variable-sized data packets that are issued at periodic intervals. This would the case, for example, for MPEG (Moving Pictures Experts Group) video transmission. The base station provides periodic unicast requests that are in line with the flow's real-time needs and allow the subscriber stations to specify the size of the grant required. This service requires more resources than UGS because of the request overheads it requires but supports grant sizes of different sizes for optimum efficiency in real-time data transport[6].

### C. Extended Real-time Polling Service (ertPS)

The ertPS is a scheduling mechanism that builds on the efficiency of both UGS and rtPS. UGS allocations are fixed in size, ertPS allocations are dynamic. The ertPS is suitable for variable-rate real-time applications that have data rate and delay requirements. Priority for packets is Normal [7].

### D. Non-real-time Polling Service (nrtPS)

The nrtPS is designed to support delay-tolerant data streams consisting of variable-size data packets for which a minimum data rate is required. The standard considers that considered that this would be the case, for example, for an FTP transmission. The base station provides regular unicast uplink requests to guarantee request opportunities even during network congestion. A CID is a unique connection identifier assigned to every connection through the base station. *nrtPS* states that base station polls nrtPS CIDs at every second or less[8].

### E. Best Effort(BE)

The BE service is designed to support data streams for which no minimum service guarantees are required and therefore may be handled on *bes*t basis.

## 2. RELATED WORK

The authors in [9 presented the use of two differing QoS level schemes; Best Effort level and High Effort level for the development of streams. Their proposed framework involves a center point that isolates streams with different level needs by dispensing them towards a similar goal. This was done using a procedure with two tables DRT (Dedicated Routing Table) and Standard

Routing Table (SRT) responsible for the movement of data streams. They proposed QOLSR (Quality Optimized Link State Routing) as the basic OLSR protocol was considered for development. This proposal did not consider extremely important aspects of QoS information exchange like reservation signaling, Connection Admission Control, and the stream classifier

Authors in [9] proposed scheduling strategies for nrtPS connections in IEEE 802.16 networks. The three strategies were categorized into dynamic,cross-layer, and conventional schedulers. The authors, in their review paper, additionally explored neuro-fuzzy scheduling and game theory-based techniques for further optimization techniques. They also highlighted that theories around the development of nrtPS are still at their growing stages and the scope is huge. They also suggested that one way to improve the *nrtPS* traffic class would be to use a technique that bases its scheduling decision based on the queue length of the nrtPS class. This would be through the determination of bandwidth requests by nrtPS and storing these requests in a virtual queue at the base station. The bandwidth requests are then sorted by the lowest queue length and each virtual queue is assigned a counter. The algorithm then does verification of the bandwidth requests made by the connection if whether they were satisfied or not. If the connection is not satisfied, then the algorithm checks the availability of more symbols to be allocated. If there are more symbols, then they are allocated to the connection and its counter will be decreased. The proposed algorithm tries to evade starvation of *nrtPS* connections. The proposed algorithm was intended for non-real-time applications and that excludes many high performing scheduling classes like UGS and ertPS

Authors in [10] presented a QoS aware framework to bridge the gap between security and QoS for the optimal functioning of a MANET. The study presented the existing challenges, attacks, and architectures as highlighted in the literature. They also presented a security keying system linked with the basic configuration of OLSR. Although this work used UGS as their preferred QoS scheduling class, the focus was on security and QoS attainment. The results with UGS showed better QoS than other scenarios.

Authors in [11] proposed a queue length scheduling algorithm for non-real-time and real-time traffic. The scheduling is initiated based on the number of MPDUs present at the start of the uplink subframe. The algorithm was designed to help in providing excess resources to non-real-time packets and also considering the queue length while guaranteeing QoS. The algorithm was not designed for real-time applications like video conferencing.

The authors in [12] proposed a method to improve the protocol Dual Busy Tone Multiple Access(DBTMA) using two elements called: busy tones and Ready To Send/Clear to Send (RTS/CTS) dialogues. A strategy to improve the effectiveness of fast retransmission was used. This retransmission strategy involved using a negative acknowledgment after a collision is caused by hidden nodes. The hidden node then listens to the negative acknowledgment signal to determine the requirement fast retransmission scheme. This method was compared with other methods in terms of their various network parameters and it showed improved QoS in terms of throughput, packet delivery ratio, and packet loss as compared to existing architectures but did not specifically speak to QoS scheduling.

Authors in [13] proposed a Medium Access Control (MAC) protocol that defines mechanisms for QoS provision and bandwidth allocation. The authors did not, however, specify the details on how to schedule traffic. This is necessary so that the vendor can differentiate their product through implementation. The authors introduce an efficient QoS scheduling strategy based on the distributed and hierarchical architecture for IEEE 802.16 systems. The simulation results provided positive feedback in terms of QoS guarantees for all types of traffic as defined by the IEEE 802.16 standard.

## 3. MATERIALS & METHODS

The research methodology used in this work is simulation as according to [14], simulation is defined as an experimentation method which involves the creation of an artificial environment within which relevant data can be generated. This is done in a controlled environment and it permits the observation of system dynamic behaviors. One of the purposes of simulation is to perform real decision making or diagnostic tasks. Decisions taken through analytical formulations are less preferred compared to simulations because they do not provide certainty. Network simulators best describe and represent the state of the network. These include nodes, links, switches, hubs, and routers. In modern-day simulators, they are either Graphical User Interface (GUI) driven or Command Line Interface (CLI) driven. Simulation is mainly used in performance analysis, comparison, or even management and also for determining how a network would behave in a real-life situation. The generated result of the simulation helps in identifying the performance attributes of the network. The flair to the entire simulation software is that there is a wide range of tools that ensure the generation of excellent results (GUI-based). There is a choice of network traffic selection, programming environments, projection, and statistical data representation. These are all part of the package of the simulation tool.

Discrete event simulation and experimentation using NetSim are implemented to adequately characterize variables, corresponding states, and events that change the value of these variable states in some rule-oriented but stochastic manner. The entities are the different components in the proposed framework.

In this study, the QoS measures used to ensure that quality is optimized in the network are throughput, delay, and jitter:

**Throughput:** It represents the number of bits forwarded from the Medium Access Control (MAC) to higher layers in all nodes in the network. It is measured in bits per second. The throughput may also be referred to as the average number of packets successfully transmitted or received per second. This work focused on the application throughput which is the total user data sent to the intended destination per second as shown in equation 1.

$$\text{Throughput} = \frac{P_d}{t} \quad (1)$$

Where $P_d$ is the number of packets delivered and $t$ is the time in seconds.

**Delay:** This is normally the time taken for one packet to be transmitted from the source node to the destination node. This performance metric evaluates the routing protocol's effectiveness in the use of network resources. Delay may be caused by several obstacles including transfer time, buffering during discovery latency, interference queue, and propagation and it is represented as shown in equation 2:

$$\text{Delay} = T_{rx} - T_{st} \quad (2)$$

Where $T_{rx}$ is the time the packet is received and $T_{st}$ is the time the packet is sent.

**Jitter:** This is the variation in time between route changes and data packets arriving. The variation may be caused by internal sources like data transmission errors, the presence of a malicious node, and network congestion. It usually affects the audio quality of the video if its level is high. The formula to calculate jitter is shown in equation 3:

$$\text{Jitter} = D_t - D_p \qquad (3)$$

Where D*t* is the transmission delay of the current packet and D*p* transmission delay of the previous packet.

Table 1: Simulation parameters of the Network

| Parameters | Value(s) |
|---|---|
| Simulator | NetSim Standard v12.1 |
| Application Protocols | OLSR |
| Grid length | 1000m*1000m |
| Simulation time | 100seconds |
| Traffic type | Video conferencing |
| QoS Class | rtPS,ertPS,BE,nrtPS & UGS |
| Node movement model | Random Waypoint |
| Trajectory | Random |
| Transport Protocol | UDP |
| Speed | 50km/h |
| Refresh interval | 2s |
| Encryption Algorithm | Advanced Encryption Standard |
| Node density | 10 |

## 4. RESULTS & DISCUSSION

The simulation results for the QoS running video-conferencing application for 10 nodes according to throughput, delay, and jitter:

### 4.1. Throughput

*A. Best Effort, ertPS, NrtPS , rtPS & UGS*

For Best Effort, the first 1000ms show a sharp increase in throughput levels up to approximately 0.35Mbps then continues to a nearly constant figure for the rest of the simulation time up until it reaches a peak of about 0.447Mbps after 7800ms. *Best effort* is considered low priority in terms of scheduling but gave an impressive throughput level for video streaming. For *ertPS*, the first 2000ms represents a refresh interval time before the application started. The next 18000ms shows a sharp rise in throughput levels up to 0.37Mbps as shown in fig.2. The highest recorded throughput is 0.458 at the end of the simulation time.

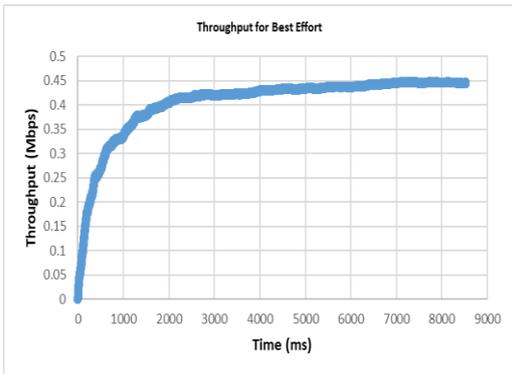
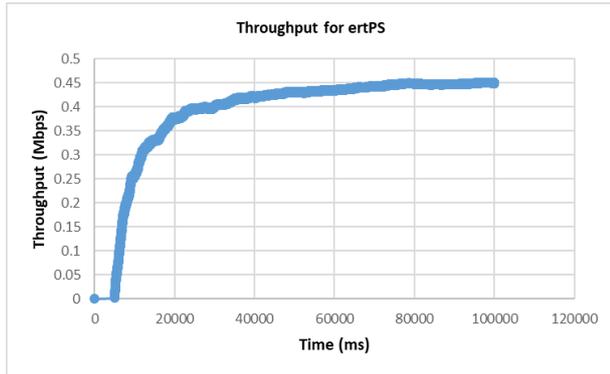

Fig.1: Throughput for Best Effort    Fig.2: Throughput for ertPS

For *nrtPS*, as shown in Fig.3, the first 2000ms represents a refresh interval time before the application started. The next 18000ms shows a sharp rise in throughput levels up to 0.375Mbps. The highest recorded throughput is 0.45Mbps at the end of the simulation time. In fig.4, the first 2000ms represents a refresh interval time before the application started for *rtPS*. The highest recorded throughput is 0.451Mbps. In the first 18000ms of *UGS's* performance, there is a sharp rise in throughput levels up to 0.377Mbps as shown in fig.5. The highest recorded throughput is 0.45Mbps at the end of the simulation time. *UGS* is known for high priority scheduling performed averagely or rather less than expected in terms of throughput as compared to *nrtPS*. The throughput levels from all three scheduling algorithms are virtually similar.

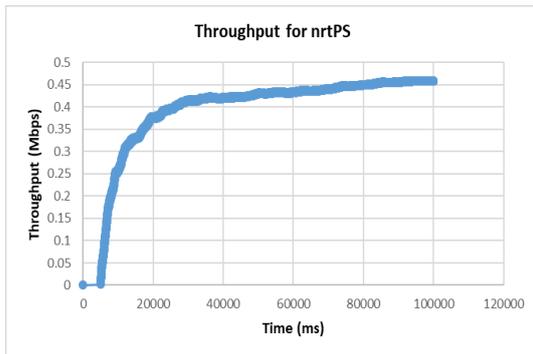
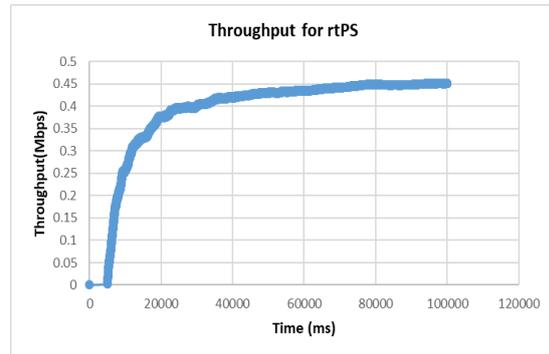

Fig.3: Throughput for nrtPS    Fig.4: Throughput for rtPS

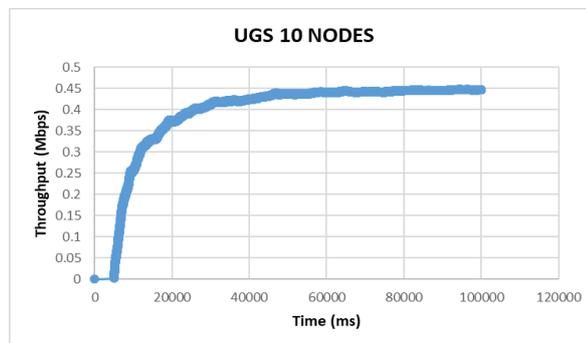

Fig. 5 : Throughput for UGS

## 4.2. Delay

According to Fig 6, the QoS classes portrayed a consistent delay for all 10 nodes. *Best Effort* obtained 6371.5 microsec and *ertPS* obtained 6262.1 microsec for all 10 nodes. *rtPS* had 6275.46microsec, ertPS obtained 6262.096 microsec and *nrtPS* obtained 6519.375 microsec for all 10 nodes. The performance of *UGS* in terms of delay is far better as compared to the other algorithms. *nrtPS* obtained the highest delay because it mostly works for non-real-time applications. *Best effort* also uses low priority scheduling for packets and this is not ideal for video streaming applications. An alternative to *UGS* would be *ertPS* and *rtPS* as they are accommodative to video streaming. The other scheduling algorithms are not weaker than the recommended ones but it depends on the application traffic that is being implemented.

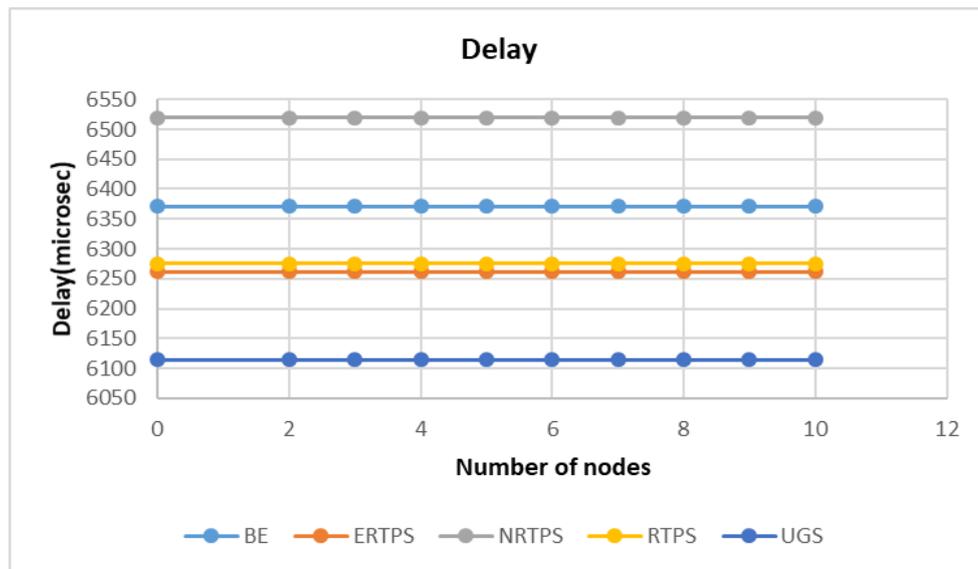

Fig.6 : Delay for all QoS classes

## 4.3. Jitter

The QoS classes have a consistent jitter level for all 10 nodes as shown in Fig 7. *Best Effort* obtained 2732.75 microsec and *ertPS* obtained 2663.96 microsec for all 10 nodes. *rtPS* had 2683.74 microsec, and *nrtPS* obtained 6519.375 microsec for all 10 nodes. The performance of *UGS* in terms of jitter is slightly higher than *ertPS*. *ertPS* builds on the efficiency of both UGS and RTP hence its excellent performance with the only difference in that its allocations are dynamic as compared to UGS whose allocations are fixed. *nrtPS* got the highest jitter because it mostly works for non-real-time applications and the application ran was in real-time. *Best effort* also uses low priority scheduling for packets and this is not ideal for video streaming applications hence its high jitter levels. An alternative to *ertPS* would be *UGS* and *rtPS* as they are accommodative to video streaming and share a few properties.

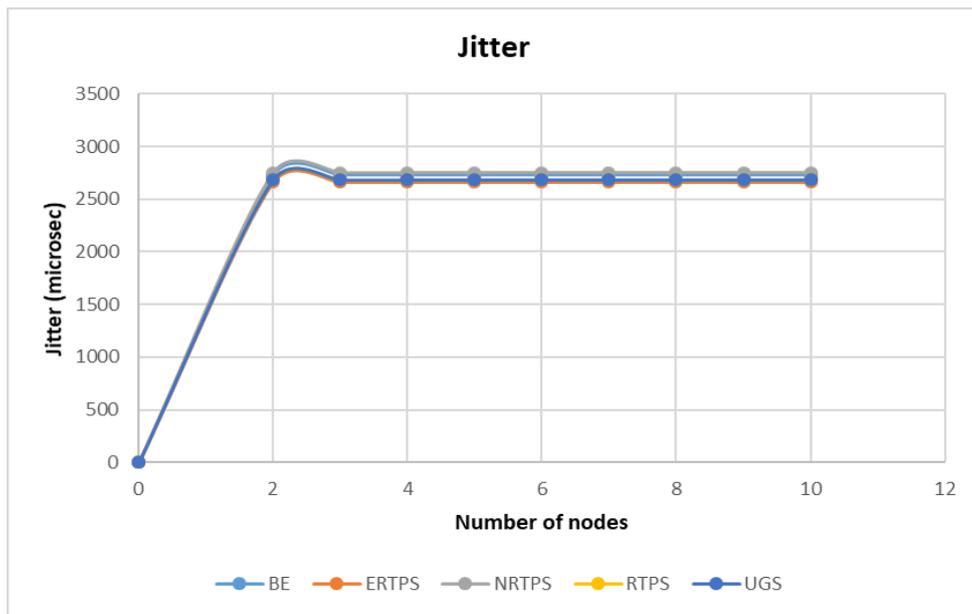

Fig.7: Jitter for all QoS classes

## 5. CONCLUSIONS

This paper presented and discussed an analysis of different QoS scheduling classes running video conferencing applications in MANETs. The comparative analysis aimed to draw a class that is best suitable for streaming purposes in a MANET in terms of QoS provision. The simulation results provided unique results in terms of the performance of these QoS classes. UGS, a high priority scheduling class, was outperformed in terms of throughput and jitter by ertPS. The only logical explanation for this unique result is that ertPS uses UGS and rtPS properties and that it capitalizes on its dynamic scheduling properties. As future work, the utmost intention is to further improve the performance of *ertPS* by analyzing the source code and making improvement of its parameters. The new source code will be an addition to improve the quality of the ertPS protocol for optimum performance. The advanced optimization techniques using ertPS protocol is another option that will also be considered.

### ACKNOWLEDGEMENTS

This work would not be possible without the support from the Faculty of Natural and Agricultural Sciences, its Postgraduate office of the North-West University and our partners at TETCOS, India who provided the test-bed for our practical and evaluation of this work.